\documentclass[a4paper]{article}

\usepackage{INTERSPEECH2020}
\usepackage{hyperref}
\usepackage{threeparttable}
\usepackage{multirow}

\title{Neural Language Modeling With Implicit Cache Pointers}
\name{Ke Li$^{1}$, Daniel Povey$^3$, Sanjeev Khudanpur$^{1,2}$\thanks{This work was partially supported by unrestricted gifts from \href{https://research.fb.com/programs/faculty-awards/page/2/?dateYear=2015}{Facebook} and \href{https://www.apptek.com/}{Applications Technology (AppTek)}.}}
\address{
  $^1$Center for Language and Speech Processing \& $^2$Human Language Technology Center of Excellence \\ 
  The Johns Hopkins University, Baltimore, MD 21218, USA.\\
  $^3$Xiaomi Corp., Beijing, China.}
\email{\texttt{\{kli26,khudanpur\}@jhu.edu}, \texttt{dpovey@gmail.com}}

\begin{document}
\ninept

\maketitle
\begin{abstract}
  A cache-inspired approach is proposed for neural language models (LMs) to improve long-range dependency and better predict rare words from long contexts. This approach is a simpler alternative to attention-based pointer mechanism that enables neural LMs to reproduce words from recent history. Without using attention and mixture structure, the method only involves appending extra tokens that represent words in history to the output layer of a neural LM and modifying training supervisions accordingly. A memory-augmentation unit is introduced to learn words that are particularly likely to repeat. We experiment with both recurrent neural network- and Transformer-based LMs. Perplexity evaluation on Penn Treebank and WikiText-2 shows the proposed model outperforms both LSTM and LSTM with attention-based pointer mechanism and is more effective on rare words. $N$-best rescoring experiments on Switchboard indicate that it benefits both very rare and frequent words. However, it is challenging for the proposed model as well as two other models with attention-based pointer mechanism to obtain good overall WER reductions.
\end{abstract}

\section{Introduction}
\label{sec:intro}

Neural language models (LMs) are an important module in automatic speech recognition (ASR) \cite{mikolov2010recurrent,chen2015recurrent,xu2018neural}.
Standard recurrent neural network language models (RNNLMs) make predictions based on a fix-sized hidden vector, making modeling long-range dependency challenging.
Although LSTMs outperform vanilla RNNs, it has been observed that they usually retain only a relatively short span of context~\cite{khandelwal2018sharp,chelba2017n}. Memory augmented models and attention mechanism have been proposed to increase the hidden state's capacity to retrieve information from hidden states in the more distant past. Though improved performance has been reported, RNNLMs with the standard softmax output still struggle with rare or unknown words, even with attention.

Since the self-attention architecture was proposed~\cite{vaswani2017attention}, deep Transformers have demonstrated state-of-the-art performance on natural language processing tasks~\cite{devlin2018bert, Radford2018ImprovingLU, dai2019transformer}. Transformer-based LMs have outperformed RNNLMs on large corpora and been used in rescoring stage in ASR systems~\cite{irie2019language,li2020empirical}. However, their ability to capture long-term dependency, e.g. self-trigger effects (word repetitions), remains unclear.

In real scenarios, especially in conversations, after a word or phrase is spoken, it is highly likely to be spoken again \cite{lau1993trigger,church2000empirical}. These self-triggers or topic-word effects can be captured by cache models, which stores the unigram distribution of recently seen words. Cache models adapt pre-trained LMs to local contexts (decoded hypotheses) in ASR systems and hence can improve ASR performance~\cite{kuhn1990cache,li2018recurrent}. Usually, cache models are integrated in pre-trained models at test time. It is a lightweight approach as no model retraining is required, while it may not be optimal. Effectively incorporating them in training stages and enabling neural LMs to learn to adapt to recent history remain to be explored.

In this work, we propose a cache-inspired approach for neural LMs to improve the capability of modeling long-term dependency, especially for rare words. The output is extended by a predefined size $L$ to represent $L$ preceding words in history. The pre-softmax activation of the $L$ units, like the other pre-softmax units, is computed by a linear transformation of the hidden state or context vector, and then appended to the output before the softmax layer, as shown in \autoref{fig:cache}. The training loss is still cross entropy. However, unlike standard training, wherein supervision comes from a vocabulary-sized one-hot vector encoding the predicted word, the supervision vector is now $L$ bits longer and contains additional ones in each history position where the word is the same as the predicted one. 

The extended output and modified supervision implicitly enable learning \emph{from where} in history \emph{to copy}. While it may still be difficult for the model to learn which words are particularly likely to be self-triggers, i.e. \emph{when to copy}. To provide a mechanism for this, one additional unit is introduced in the pre-softmax layer (but not included in the softmax computation) to capture the probability that the current word may be a self-trigger. At each word position, activations from these additional units in the $L$ previous positions are added to the $L$ extended output units.

Though cache-inspired neural LMs for improving long-range dependency have been proposed and demonstrated superior performance than LSTMs in terms of perplexity, to our best knowledge, their effect on ASR accuracy remains to be explored. In this study, we evaluate neural LMs on ASR tasks. We also apply the proposed approach to Transformer architecture to verify if cache-based information is still beneficial.

\section{Related Work}
\label{sec:label}
In this section, we briefly introduce related work about approaches to improve performance on rare words and long-term dependency for sequence modeling problems including neural language modeling and machine translation~\cite{vinyals2015pointer, gu2016incorporating, gulcehre2016pointing,merity2016pointer,grave2016improving,krause2017dynamic}. 
Vinyals et al.~\cite{vinyals2015pointer} introduces an attention-based pointer network to select items from the input as output. It has been shown to help on geometric problems~\cite{vinyals2015pointer}. The pointer network can also improve performance of text summarization~\cite{gu2016incorporating,gulcehre2016pointing} and alleviate issues of rare or unknown words in neural machine translation~\cite{gulcehre2016pointing}.  

For neural LMs, similar ideas have been proposed to better model long-range dependency~\cite{merity2016pointer,grave2016improving}. The most relevant work is the pointer sentinel mixture model (PSMM), a mixture model of a standard LSTM and an auxiliary pointer network which captures the unigram distribution of history words via attention~\cite{merity2016pointer}. The mixture weight is jointly optimized. A similar mixture model, neural cache model~\cite{grave2016improving}, differs from PSMM in aspects such as the query vector for computing attention scores is hidden state itself instead of a projected version and it does not require model retraining. 
The motivation of dynamic evaluation~\cite{krause2017dynamic} is similar to the neural cache model, but the implementation is different: it adjusts model parameters via gradient updates based on partial predicted sequences during test time. It may be viewed as a modified version of the dynamic updating method proposed by Mikolov et al.~\cite{mikolov2010recurrent}.

\section{Proposed Model}
\label{sec:methods}
Language modeling can be framed as predicting the next word (target) given preceding words (history). It usually can be observed that some words tend to be much more likely targets once they have occurred in the history. PSMM learns to ``reproduce'' a word from recent history by an attention-based pointer network. And its mixture weight is computed by a specially designed gating mechanism. Though the PSMM achieves lower perplexity than standard LSTM, the attention and gating mechanisms are relatively complex and not the only approach to do so. We aim to achieve a similar effect with simpler models.

\subsection{Pointer Component}
Let us denote the hidden state of an RNNLM at time step $t$ as $\mathbf{h}_{t}$. 
Conventionally, the RNNLM output $\mathbf{y}_{t}$ is determined as
\begin{equation}
    \mathbf{y}_{t} = \text{softmax}(\mathbf{W}\mathbf{h}_{t} + \mathbf{b}),
    \label{eq:output}
\end{equation}
where $\mathbf{W} \in \mathbb{R}^{V\times H}$, $\mathbf{b} \in \mathbb{R}^{V}$, and $\mathbf{y}_{t} \in \mathbb{R}^{V}$, with $V$ and $H$ being the vocabulary size and hidden state dimension, respectively.

\begin{figure}
    \centering
    \includegraphics[width=0.9\linewidth]{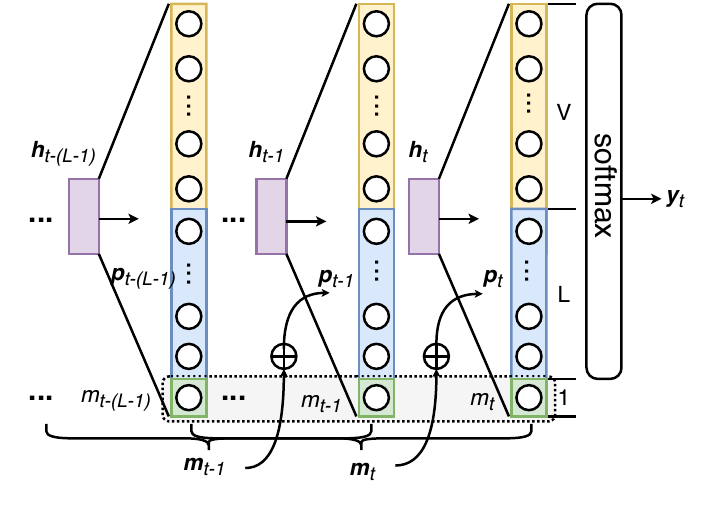}
    \caption{Neural LMs with implicit cache pointers.}
    \vspace*{-\baselineskip}
    \label{fig:cache}
\end{figure}

We extend the output dimension by a predefined size $L$. The extended part represents the $L$ immediately preceding words in history. Activation of these $L$ extended units, denoted as $\mathbf{p}_{t}$ in \autoref{fig:cache}, is computed via a linear projection of $\mathbf{h}_{t}$ from the last hidden layer of the RNNLM.
We thus have
\begin{align}
    &\mathbf{p}_{t} = \mathbf{W}_{p}\mathbf{h}_{t},\\
    \label{eq:z_t}
    &\mathbf{z}_{t} = \text{concat}(\mathbf{W}\mathbf{h}_{t} + \mathbf{b}, \mathbf{p}_{t}),\\
    \label{eqn:softmax_zt}
    &\mathbf{y}_{t} = \text{softmax} \left(\mathbf{z}_{t}\right),
\end{align}
where $\mathbf{W}_{p} \in \mathbb{R}^{L \times H}$, $\mathbf{z}_{t} \in \mathbb{R}^{V + L}$, and $\mathbf{p}_{t} \in \mathbb{R}^{L}$. Applying softmax on $\mathbf{z}_{t}$ generates the extended output $\mathbf{y}_{t} \in \mathbb{R}^{V + L}$.

Since the $L$ extended outputs indicate \emph{where} to copy \emph{from} the history, we call $\mathbf{p}_t$ the \emph{pointer} component of our model. It only introduces $L$ $\times$ $H$ additional parameters.

The objective for training a neural LM is to maximize the log likelihood of training data. The loss function is written as
\begin{equation}
    L(\theta) 
    = -\frac{1}{T} \sum_{t=1}^{T} \log( \mathbf{y}_{t} \boldsymbol{\cdot} \mathbf{s}_{t}),
\end{equation}
where $T$ is the total number of words in the training data, $\mathbf{s}_{t}$ is the supervision vector, and $\boldsymbol{\cdot}$ is vector dot product operation. 
In conventional training, $\mathbf{s}_{t}$ is a one-hot vector with 1 in the index of the target word. To train the proposed neural LM with the pointer component, the supervision vector $\mathbf{s}_{t}$ is set to have \emph{additional} ones in history positions where the the target was previously seen. So $\mathbf{s}_{t}$ is an \textit{at-least-one-hot} vector.

\subsection{Memory Augmented Pointer Component}
The pointer component and the modified training supervision makes an RNNLM be aware of where to copy from the history. However, it may still be challenging for the model to memorize which words are particularly likely to reoccur, i.e. are "bursty". To learn the burstiness of words, one additional unit, denoted by $m_{t}$ is introduced alongside the pointer $\mathbf{p}_t$, as shown in \autoref{fig:cache}.

This additional unit is computed by a vector dot product of the hidden state $\mathbf{h}_t$ and a parameter vector with the same dimension as $\mathbf{h}_t$, but is not used in computing the softmax. 
It influences the probability that a word may repeat through $\mathbf{p}_t$.  Specifically, these additional units from the $L$ immediately preceding word positions are concatenated to form
\begin{equation}
\mathbf{m}_{t} = \text{concat}(m_{t-(L-1)},...,m_{t}),
\end{equation}
where $\mathbf{m}_{t} \in \mathbb{R}^{L}$, and $\mathbf{m}_{t}$ is element-wisely added to the pointer component $\mathbf{p}_{t}$, i.e. 
\begin{equation}
    \mathbf{p}_{t} := \mathbf{p}_t + \mathbf{m}_{t}.
    \label{eqn:mem-augmentation}
\end{equation}
%
Thus $m_t$ influences the output $\mathbf{y}_{t}$ indirectly by modifying $\mathbf{p}_t$ in (\ref{eq:z_t}), which in turn is a part of $\mathbf{z}_t$ in (\ref{eqn:softmax_zt}).

Compared with an RNNLM, the memory augmented pointer component only has $(L+1)$ $\times$ $H$ total additional parameters while a PSMM has extra $H^{2} + 2 H$ parameters. Without attention and gating mechanism, the proposed model is simpler than PSMM and has fewer extra parameters when $L$ $\le$ $H$.  
%
%

This pointer mechanism described above is for RNNLMs, while it can also be easily incorporated into Transformer-based LMs. In the latter, the context vector from the last Transformer block is treated as the hidden state in RNNLMs. 

\section{Experimental Setup}
We conduct experiments on two text datasets, Penn Treebank (PTB) and WikiText-2~\cite{merity2016pointer}, and two ASR corpora, Switchboard (SWBD) and Wall Street Journal (WSJ). We use Kaldi RNNLM~\cite{xu2018neural} for data preprocessing on SWBD, e.g. including the English Fisher corpus, and WSJ. Sentences in SWBD+Fisher interleave conversation turns, as derived from time-information in transcriptions. Statistics of the datasets are shown in~\autoref{tab:datasets} (``sent len'' is average sentence length).
\begin{table}[ht]
  \caption{Statistics of datasets used in experiments.}
  \label{tab:datasets}
  \centering
  \scalebox{0.85}{
  \begin{threeparttable}
  \setlength\tabcolsep{2pt}
    \begin{tabular}{ c | c| c |c| c | c }
    \toprule
    Dataset & \#\,words\tnote{*} \space & $|$Vocab$|$ & sent len & OOV (train\,/\,dev\,/\,test) & Style \\
    \midrule 
    PTB & 929K & 10K & 21 & 4.8\%\,/\,4.7\%\,/\,5.8\% & written\\
    WikiText-2 & 2M & 33K & 22 & 2.6\%\,/\,5.4\%\,/\,6.2\% & written\\
    \midrule
    SWBD+Fisher & 34M & 30K & 10 & 8.9\%\,/\,0.0\%\,/\,5.8\% & spoken\\
    WSJ & 39M & 123K & 23 & 4.6\%\,/\,4.6\%\,/\,5.6\% & written\\
    \bottomrule
    \end{tabular}
    \begin{tablenotes}
        \footnotesize
        \item[*]{The end of sentence token is included in the count of training words.}
    \end{tablenotes}
  \end{threeparttable}}
  \vspace*{-\baselineskip}
\end{table}

We develop baselines with both LSTM and Transformer-based LMs. Model details are present in \autoref{tab:models}. Plain LSTMs are baselines from each dataset except for WSJ. We has a stronger baseline for PTB and WikiText-2: AWD-LSTM~\cite{merity2017regularizing} with frequency-agnostic word embeddings~\cite{gong2018frage}, denoted by Frage-AWD-LSTM.
For SWBD+Fisher, the stronger baseline is a Transformer LM with self-attention~\cite{vaswani2017attention}. We only experiment with Transformer architecture on WSJ.
Given our academic computational resources, we were unable to make comparisons with even stronger baselines, e.g. GPT and BERT, or with the optimized architectures of~\cite{wang2019cas}, which requires industrial-strength resources. 

All neural LMs are on word level, implemented with Pytorch, and optimized via SGD\footnote{For the Transformer LMs, we tried Adam with the learning rate schedule proposed in~\cite{vaswani2017attention}, but failed to get better performance than SGD.}. We tie the embedding and output matrices in all setups. The dropout rate for PTB and WikiText-2 is 0.5, while for SWBD+Fisher it is 0.1 for both LSTM and Transformer LMs. Parameters of Frage-AWD-LSTMs not listed in~\autoref{tab:models} follow the settings in~\cite{gong2018frage}. 

\begin{table}[ht]
  \caption{Details of neural network dimensions for various LMs.}
  \label{tab:models}
  \centering
  \scalebox{0.85}{
  \begin{threeparttable}
  \setlength\tabcolsep{1.5pt}
    \begin{tabular}{ c | c | c | c | c}
    \toprule
    Model & Corpus & Layers & Units & Heads \\
    \midrule
    Plain LSTM& All (except for WSJ) & 2 & 650 & - \\
    \midrule
    Frage-AWD-LSTM & PTB/WikiText-2 & 3 & 1150 & -\\
    \midrule
    Transformer & SWBD+Fisher/WSJ & 6 & 512/768 & 8\\
    \bottomrule
    \end{tabular}
  \end{threeparttable}}
\end{table}

For ASR experiments on SWBD and WSJ, we use the Kaldi toolkit~\cite{povey2011kaldi} to train acoustic models and perform $N$-best rescoring. Acoustic models are factorized TDNNs~\cite{povey2018semi}, trained using the LF-MMI objective~\cite{povey2016purely}. We do not include Fisher audio to train acoustic models for SWBD. To rescore each of the $N$ hypotheses for an utterance, we find it useful to initialize the initial LM state 
with the last LM state of the best hypothesis for the previous utterance.

\section{Experiments}
\label{sec:exps}
\subsection{Perplexities on PTB and WikiText-2}
\label{ptb-wiki-ppl}
We first compare the proposed model with PSMM and neural cache under the plain LSTM setup. Perplexities on PTB and WikiText-2 are shown in Table~\ref{tab:ptb-wiki2}. 
The performance gap between the PSMM in the paper~\cite{merity2016pointer} and ours is mainly caused by different implementations of truncated back-propagation through time (BPTT). They use an explicit truncated BPTT while we follow the normal way discussed in~\cite{merity2016pointer} considering efficiency and convenience of data preprocessing. We first concatenate all text words and then chunk them with fixed size $L$. So, if the truncated BPTT length is $L$, each training word on average experiences $L$/2 instead of $L$ time-steps for back-propagation. This means each training word sees $L$/2 history words on average.

\interfootnotelinepenalty=10000
\begin{table}[ht]
    \setlength{\tabcolsep}{1.0pt}
    \caption{Perplexities on PTB and WikiText-2 (plain LSTMs).}
    \label{tab:ptb-wiki2}
    \centering
    \scalebox{0.85}{
    \begin{threeparttable}
    \begin{tabular}{ l r r r | r r r }
    \toprule
        \multirow{2}{*}{Model} & \multicolumn{3}{c}{PTB} & \multicolumn{3}{c}{WikiText-2}\\
        \cmidrule(lr) {2-4} \cmidrule(lr) {5-7}
        & \#Params & Dev & Test & \#Params & Dev  & Test \\
        \midrule
        5gram KN \cite{mikolov2012context}  & 2M & - & 141.2 & - & - & -\\
        LSTM (medium) \cite{zaremba2014recurrent}  & 20M & 86.2 & 82.7 & - & - & -\\
        PSMM \cite{merity2016pointer}  & 21M & 72.4 & 70.9 & 47M\footnotemark & 84.8 & 80.8\\
        \midrule
        LSTM & 13.3M & 73.6 & 71.9 & 28.5M & 89.1 & 84.8\\
        PSMM (Ours) & 13.7M & 73.5 & 71.6 & 28.9M & 86.8 & 82.8\\
        LSTM + Neural Cache (L=50) & 13.7M & 69.3 & 68.5 & 28.9M & 81.3 & 77.0\\
        Proposed w/o Memory Aug& 13.3M & 70.1 & 69.8& 28.5M & 80.6 & 76.7\\
        Proposed w/ Memory Aug & 13.3M & \textbf{68.1} & \textbf{67.8} & 28.5M & \textbf{78.2} & \textbf{74.3}\\
        \bottomrule
        \end{tabular}
        \footnotetext{The stated number of parameters of the PSMM in \cite{merity2016pointer} is 21M. Considering the hidden state size and vocabulary, we suspect that to be a typo.  We obtain around 47M assuming no parameter tying.}
    \end{threeparttable}}
    \vspace*{-\baselineskip}
\end{table}
In Table~\ref{tab:ptb-wiki2} ``Memory Aug'' refers to the memory augmented pointer. We set history length as 100, equal to the truncated BPTT length. Setting $L$ = 50 for neural cache is a fair comparison with others. Results on both datasets show that memory augmentation provides further improvement on top of the pointer component. And with memory augmentation, the proposed model outperforms the rest on both datasets. In subsequent tables, ``Proposed'' refers to LMs with the memory augmented pointer. 

To verify whether the proposed approach is robust, we conduct experiments on a stronger baseline Frage-AWD-LSTM setup \cite{gong2018frage}. We reproduced their results and implemented the proposed approach on top of theirs, without tuning meta-parameters. Perplexity results in Table~\ref{tab:ptb-awd-lstm} shows that the proposed model on both datasets achieves better results than Frage-AWD-LSTM. Further improvements are observed with increasing the history length from 50 to 100, as expected. We also observe complementary effects of the proposed model and neural cache model.
\vspace{-2mm}
\begin{table}[ht]
    \setlength{\tabcolsep}{2.0pt}
    \caption{Perplexities on PTB and WikiText-2 (Frage-AWD-LSTM setup).}
    \label{tab:ptb-awd-lstm}
    \centering
    \scalebox{0.85}{
    \begin{threeparttable}
    \begin{tabular}{ p{4.3cm} c p{0.6cm} p{0.6cm} | c p{0.6cm} p{0.6cm}}
    \toprule
        \multirow{2}{*}{Model} & \multicolumn{3}{c}{PTB} & \multicolumn{3}{c}{WikiText-2}\\
        \cmidrule(lr) {2-4} \cmidrule(lr) {5-7}
        & \#Params & Dev & Test & \#Params & Dev  & Test \\
        \midrule
        Frage-AWD-LSTM~\cite{gong2018frage} & 24M & 58.1 & 56.1 & 33M & 66.5 & 63.4 \\
        \midrule
        Frage-AWD-LSTM (Ours) & 24M & 57.7 & 55.3 & 33M & 64.6 & 62.1\\
        Proposed (L = 50) & 24M & 55.2 & 53.9 & 33M & 60.7& 58.5\\
        Proposed (L = 100) & 24M &  54.2 &53.5 & 33M & 60.2& 57.5 \\
        Proposed + Neural Cache(L=100) & 24M & \textbf{53.0} & \textbf{52.5} & 33M & \textbf{58.0} & \textbf{55.6} \\
        \bottomrule
        \end{tabular}
    \end{threeparttable}}
    \vspace*{-\baselineskip}
\end{table}
\vspace{-2mm}

\subsection{Perplexities on SWBD and WSJ}
We experiment with both LSTM- and Transformer-based LMs on SWBD. Perplexity on dev set from Kaldi RNNLM is 50. The history length as well as BPTT length is set to 100 for the proposed model and PSMM. Results of Pytorch trained models are in Table~\ref{tab:swbd-ppls}. 
\begin{table}[ht]
    \setlength{\tabcolsep}{2.0pt}
    \caption{Perplexities on SWBD.}
    \label{tab:swbd-ppls}
    \centering
    \scalebox{0.9}{
    \begin{threeparttable}
        \begin{tabular}{ l c c c}
        \toprule
            Model & \#Params & Dev & Eval'00 \\
        \midrule
            LSTM & 26.5M & 47.0 & 	41.7 \\
            PSMM & 26.9M & 45.6 &	39.5\\
            LSTM + Neural Cache (L=50) & 26.5M & 45.1 & 39.6 \\
            LSTM + Neural Cache (L=100) & 26.5M & \textbf{44.6} & \textbf{39.1} \\
            LSTM + Proposed & 26.5M & 45.9 & 40.4 \\
        \midrule
            Transformer w/o positional embedding~\cite{vaswani2017attention}& 25.0M & 51.6 & 44.4 \\
            Transformer with positional embedding & 25.1M & 46.8 & 41.5 \\
            Transformer + Proposed & 25.1M & 45.0 & 40.2 \\
        \bottomrule
        \end{tabular}
    \end{threeparttable}}
\end{table}
For LSTM-based models, the proposed approach outperforms the baseline LSTM, but performs slightly worse than the PSMM and neural cache models.
For Transformer LMs, the proposed approach also achieves better perplexity than the two Transformer baselines. 

We notice the performance gains of the proposed model over both LSTM and Transformer baselines on SWBD are smaller than on PTB and WikiText-2. To check whether this may relate to style (only SWBD is spoken style) and average sentence length (sentences on Switchboard are the shortest on average), we experiment with Transformer-based LMs on WSJ. The proposed approach reduces perplexity from 71.5 to 65.6 on test set (eval92), compared with a baseline Transformer LM. The improvement is relative 8.2\% and in similar range with gains on WikiText-2. And this results in 0.1 absolute WER reduction from 1.5 to 1.4 on the same test set by $N$-best rescoring. While no conclusions can be drawn yet, experiments show that the proposed model perform better on written-style text which usually has longer average sentence length. 

\vspace{-2mm}
\subsection{Analysis of Impact on Rare Words}
\label{sec:rare-analysis}
\begin{figure}[h]
    \centering
    \includegraphics[width=0.9\linewidth]{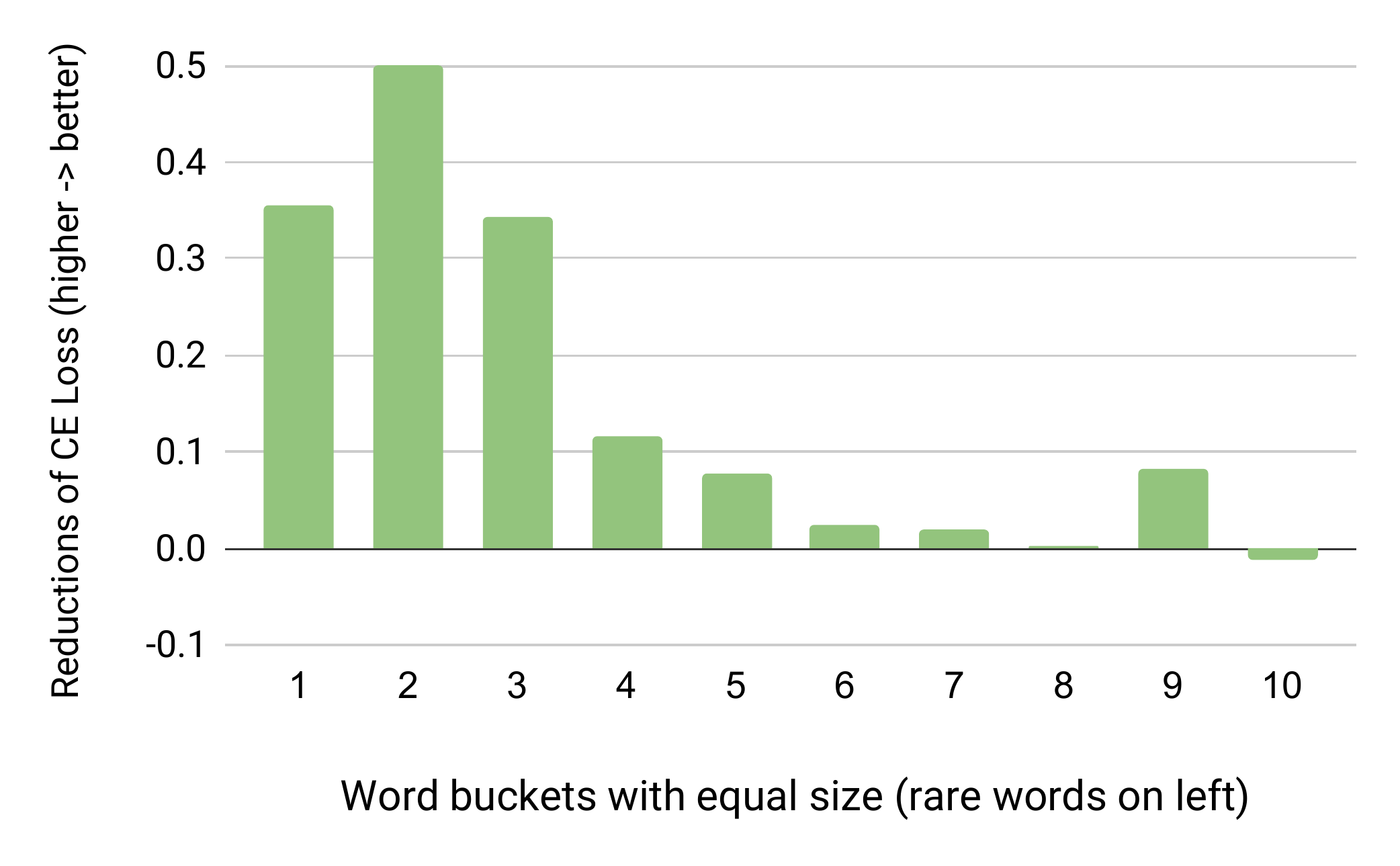}
    \caption{Cross-entropy reduction from the proposed model w.r.t an LSTM on WikiText-2.}
    \vspace*{-\baselineskip}
    \label{fig:analysis_wikitext2}
\end{figure}

One desired feature of the proposed model is that it may predict rare words better than LSTMs. To verify this, we further examine LM performance on the test set of each dataset. We split the vocabulary of each corpus into 10 buckets based on word frequencies in training data such that we get a roughly equal number of test tokens in each bucket. We then compute the differences between the test cross entropy of the proposed model and the LSTM baseline on words in each bucket for each dataset. 
\autoref{fig:analysis_wikitext2} shows the results on WikiText-2. As expected, larger reductions in cross-entropy are observed from the proposed model on rare words. Similar trends are seen on PTB and SWBD, though the overall perplexity improvement on SWBD is marginal. Figures for them are omitted due to page limits. 

\vspace{-2mm}

\vspace{-2mm}
\subsection{Rescoring Evaluation and Analysis on SWBD}
As Transformers show similar perplexity gains as LSTMs on SWBD, we only experiment with LSTMs for $N$-best rescoring. WERs on the full HUB5'00 evaluation set (Eval{'}00), the SWBD subset (SWB), and Callhome subset (CH) are in Table~\ref{tab:swbd}. ``State-carry'' means when scoring a hypothesis for current utterance, the initial hidden state is copied from the last hidden state of the best hypothesis for the previous utterance, instead of being zero initialized. 
WER improvements by the LSTM with ``state-carry'' in Table~\ref{tab:swbd} indicate that cross-sentence context is useful. Similar observations are presented in~\cite{irie2019training}. If not specified with ``w/o state-carry'', models are evaluated in the state-carry way. 

\begin{table}[h]
    \setlength{\tabcolsep}{2.0pt}
    \caption{WERs by N-best rescoring with baselines and the proposed model on SWBD.}
    \label{tab:swbd}
    \centering
    \scalebox{0.9}{
    \begin{threeparttable}
        \begin{tabular}{ l c c c c c}
        \toprule
            Model & Eval'00 & SWB & CH\\
        \midrule
            Kaldi RNNLM (w/o state-carry) & 11.3	& 7.5& 15.0\\
        \midrule
            LSTM (w/o state-carry)& 11.2 & 7.3 & 15.1\\
            LSTM &	10.9 & 7.1 & 14.5 \\
            PSMM & 10.9 & 7.1 & 14.6 \\
            LSTM + Neural Cache & 10.9 & 7.2 & 14.5\\
            LSTM + Proposed & \textbf{10.8} & \textbf{7.1} & \textbf{14.4} \\
            
        \bottomrule
        \end{tabular}
    \end{threeparttable}}
    \vspace*{-\baselineskip}
\end{table}

\begin{figure}[ht]
    \centering
    \includegraphics[width=0.8\linewidth]{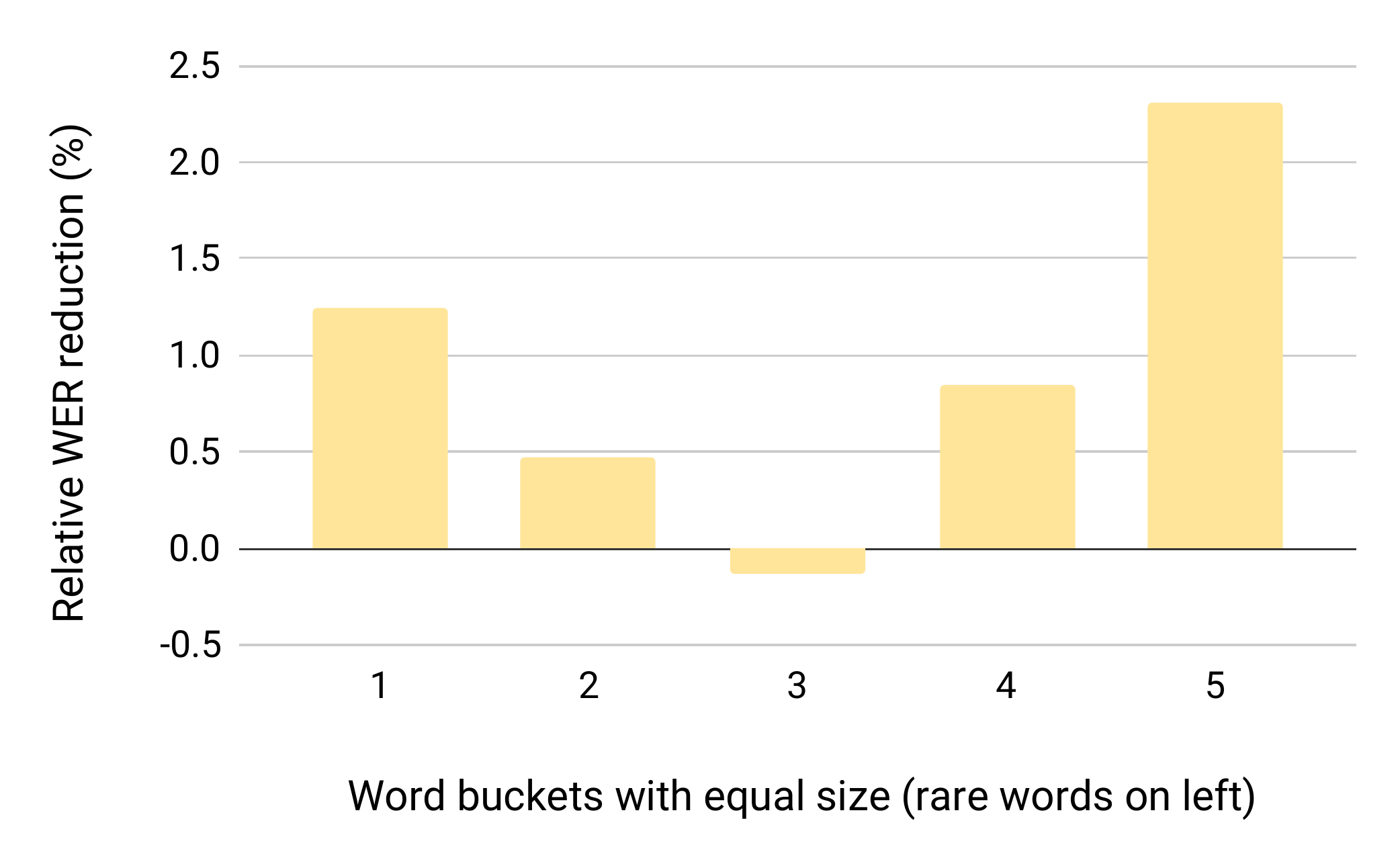}
    \caption{Relative WER reduction on by the proposed model w.r.t an LSTM on SWBD.}
    \vspace*{-\baselineskip}
    \label{fig:analysis_swbd_wer}
\end{figure}

To investigate the effect on WERs of rare words, we conduct a similar analysis as Section~\ref{sec:rare-analysis} does. Words with errors on Eval{'}00 (\textless 5000) are divided into 5 buckets based on frequency. Relative WER reductions on words in these buckets in Figure~\ref{fig:analysis_swbd_wer} show that the proposed model improves performance on both relatively rare words and very frequent ones. As expected, some rare words occur within the context window, for example, \textit{masters} and \textit{offered}, are recognized correctly. Decoded output also shows that frequent words such as \textit{train}, \textit{short}, and \textit{were} are correctly recognized. 
Though the overall WER improvement by the proposed model is marginal, correctly recognizing relatively rare words plays an important role in user experience of ASR-based products or service. 

We also notice the proposed model sometimes introduces errors on words that are wrongly recognized in first pass decoding. A possible reason is that the supervision vectors for pointer components are from decoded hypotheses and hence may contain errors. We verify this by using test transcription in rescoring and observe a further 0.1 absolute WER reduction on Eval'00 of SWBD. The mismatched condition between training and evaluation is a common issue for the proposed approach, PSMM, and neural cache model. To alleviate the mismatch, word level confidence scores and error adaptive training approaches could be considered. 

\section{Conclusion and Future Work}
\label{sec:summary}
In this work, we propose a cache-inspired pointer mechanism for neural LMs to improve the capacity of modeling long-range dependency and better predict rare words. It can be applied to both RNN- and Transformer-based models. Perplexity evaluation show that the proposed approach generally outperforms LSTM and PSMM and is more effective on rare words. Rescoring with the proposed model on SWBD and WSJ gives marginal WER improvements. Analysis shows that the mismatch between training and rescoring conditions (i.e. potentially incorrect histories) may make it challenging for both the proposed model and models with attention-based pointer network to achieve large overall WER reductions. Future work is therefore focused on methods that can mitigate the mismatch issue.

\bibliographystyle{IEEEtran}

\bibliography{mybib}


\end{document}